\newcommand{\al}{\mbox{$^{26}$\hspace{-0.2em}Al}}
\newcommand{\Msol}{M_{\sun}}

\newcommand{\gray}{\mbox{$\gamma$-ray}}

\newcommand{\pcmq}{\mbox{cm$^{-2}$}}

\newcommand{\psec}{\mbox{s$^{-1}$}}

\newcommand{\funit}{\mbox{ph \pcmq \psec}}

\def\MeV{\mbox{Me\hspace{-0.1em}V}}
\def\MeVn{\mbox{Me\hspace{-0.1em}V\hspace{-0.1em}/n}}

\def\sun{\hbox{$\odot$}}

\def   \ni {\noindent}

\def   \ssk {\vskip  5truept}

\def   \bsk {\vskip 15truept}

\def   \newpage {\vfill\eject}
\def   \newline {\hfil\break}

\documentstyle[epsfig]{article}
\begin{document}
\hsize 5truein
\vsize 8truein
\font\abstract=cmr8
\font\keywords=cmr8
\font\caption=cmr8
\font\references=cmr8
\font\text=cmr10
\font\affiliation=cmssi10
\font\author=cmss10
\font\mc=cmss8
\font\title=cmssbx10 scaled\magstep2
\font\alcit=cmti7 scaled\magstephalf
\font\alcin=cmr6
\font\ita=cmti8
\font\mma=cmr8
\def\ref{\par\noindent\hangindent 15pt}
\null

\title{\ni Gamma-Ray Line Emission from Superbubbles}

\bsk \bsk
\author{\ni Etienne Parizot$^{1}$ and J\"urgen Kn\"odlseder$^{2}$}
\bsk
\affiliation{$^{1}$Service d'astrophysique,
CEA/DSM/DAPNIA, CEA-Saclay, 91191 Gif-sur-Yvette, France}
\affiliation{$^{2}$Centre d'Etude Spatiale des Rayonnements,
CNRS/UPS, B.P.~4346, 31028 Toulouse Cedex 4, France}
\bsk
\baselineskip = 12pt

\abstract{ABSTRACT \ni
We present an evolutionary model for \gray\ line emission from
superbubbles based on shock acceleration of metal-rich stellar ejecta.
Application to the Orion OB1 association shows that \gray\ lines at
the detection threshold of the SPI spectrometer aboard of INTEGRAL
are expected, making this region an interesting target for studies of
the interaction of supernova shocks with the interstellar medium.
}
\bsk
\baselineskip = 12pt

\bsk
\baselineskip = 12pt


\text{\ni 1. INTRODUCTION
\ssk
\ni

The claim for a detection by COMPTEL (Bloemen et al.~1994) of an
intense flux of 3-7~\MeV\ gamma-rays from the Orion molecular complex,
attributed most naturally to $^{12}$C and $^{16}$O de-excitation lines,
has led many authors to re-consider the nature and impact of energetic
particles (EPs) in the interstellar medium (ISM).
Although re-analysis of COMPTEL data suggests now that the observed
emission was an instrumental artefact (Bloemen, these proceedings),
the former ``detection'' raised the question about the possible existence of
a low-energy, C and O enriched cosmic-ray component.
Indeed, independent of the COMPTEL result, new observations relating to
the Be and B abundances in the early Galaxy support the existence of
such a component (e.g.~Gilmore et al. 1992; Duncan et al.~1992;
Cass\'e et al.~1995).
A mechanism revived to explain low-energy C and O rich cosmic-rays has
been the acceleration of particles in a superbubble resulting from the
intense energetic activity of an OB association inside a molecular cloud
(Bykov \& Bloemen 1994; Parizot 1998).
Strong stellar winds and supernova (SN) explosions fill the
superbubble with both energy and enriched material to be accelerated
by the numerous secondary shocks and by magnetic turbulence
resulting from the interaction of shock waves (from winds and SNe) with
each other and with dense clumps inside the bubble
(Bykov \& Toptygin 1990; Bykov \& Fleishman 1992).
The resulting energy spectrum is expected to be very hard ($\propto E^{-1}$)
up to a cut-off energy, $E_{0}$, of $\sim100$ \MeVn.
As for the chemical composition of the EPs, it is clearly related to the
composition of stellar winds and SN ejecta, although some contamination
by swept-up and/or evaporated material is likely to occur.

In this paper we calculate the \gray\ line emission associated with
such a sce-
\newpage \noindent
nario.  As we believe that the Orion complex associated with
the Orion-Eridanus superbubble represents the most favoured target for
a detection, we normalise our results to the distance of Orion (450~pc)
and the stellar content as inferred from observations of the Orion
OB1 association (Brown et al.~1994).

\bsk
\ni 2. BASIC INGREDIENTS OF THE MODEL
\ssk
\ni

The first step in our model calculation consists in the evaluation of
the enrichment of the superbubble by stellar winds and SN ejecta as a
function of bubble age.
For this purpose we follow the evolution of a coevally formed OB
association, characterised by an IMF of slope $\Gamma$.
The enrichment is calculated using the stellar yield compilation of
Portinari et al.~(1998) who combined the Padova stellar evolutionary
models with SN models of Woosley \& Weaver (1995).
Additionally, yields for the production of radioactive \al\ have been
taken from Meynet et al.~(1997), Woosley \& Weaver (1995), and
Woosley et al.~(1995).
To determine the parameters of the superbubble ``blown'' by the
association, we derive the time dependent mechanical luminosity of the
OB association from the evolutionary tracks of the Padova group.
Using this luminosity, we solve the dynamical equation for a
spherical, homogeneous bubble (e.g.~Shull \& Saken 1995).
The characteristic density and temperature of the bubble interior is
dominated by the ``mass loading'' from evaporated gas off the shell.
This mass loading dilutes the bubble interior with ambient ISM
material which we assume to have solar composition.
We calculate the conductive mass evaporation from the shell into the
bubble by solving the equation of classical, unsaturated conductivity
(e.g.~Shull \& Saken 1995).
Even if we disposed of reliable stellar evolutionary tracks giving the
composition of the winds and the SN ejecta, we would still have to
evaluate the mixing of the ejecta with the evaporated ISM.
To avoid such a hazardous attempt, we consider two extreme scenarios, in
which the EPs are made of the stellar ejecta alone (models P), or
a perfect mixture of the ejecta and the evaporated ambient material
(models D).

The second step of our calculation consists of accelerating the enriched
material within the superbubble assuming a constant acceleration rate
during a time $\tau_{0}$ following each SN explosion.
The EP spectrum is thus normalised so that the energy injection rate
$\dot{E}$ is equal to $E_{\mathrm{SN}}/\tau_{0}$, where
$E_{\mathrm{SN}}\equiv 10^{51}$~erg is the SN energy.
To calculate the time scale $\tau_{0}$, we assume that each new supernova
influences and provides energy to a region of size $L$ around its
explosion site, in which particles are accelerated with an
efficiency $\eta \sim 10^{-3}$ (Bykov and Fleishman 1992; Parizot 1998).
Further assuming that the extension of the region in which particles
are accelerated increases as $L = v_{\mathrm{A}}t$, where
$v_{\mathrm{A}} \simeq 200$~km/s is the Alfv\'en velocity, we find
that the total energy injected in the form of EPs after time
$t=\tau_{0}$ is
$E_{\mathrm{EP}} = \eta n_{\mathrm{b}} \frac{4}{3}\pi v_{\mathrm{A}}^{3}
 \tau_{0}^{3} \langle E\rangle,$
where $\langle E\rangle$ is the mean EP energy, averaged over the
assumed spectrum, and $n_{\mathrm{b}}$ is the density of the
superbubble interior.
Equating $E_{\mathrm{EP}}$ to $E_{\mathrm{SN}}$, we obtain an estimate
for $\tau_{0}$, which we then use to normalise the EP spectrum and thus
the \gray\ fluxes.
For typical values of $\langle E\rangle = 100$ \MeVn\ and $n_{\mathrm{b}} =
10^{-2}\,\mathrm{cm}^{-3}$, we obtain $\tau_{0}\sim 10^{5}$~years,
corresponding to an acceleration power of $\sim 3\,10^{38}$~erg/s.
As argued by Bykov and Fleishman (1992), the energy spectrum of the
EPs depends on their feedback over the magnetic turbulence and the
shock waves system inside the bubble.
Any detailed calculation of this spectrum being out of the
scope of this paper, we consider here the cut-off energy,
$E_{0}$, as a free parameter with values in the range
$3-3000$ \MeVn.

\bsk
\ni 3. APPLICATION TO ORION AND DISCUSSION
\ssk
\ni

\begin{figure}
\centerline{\psfig{file=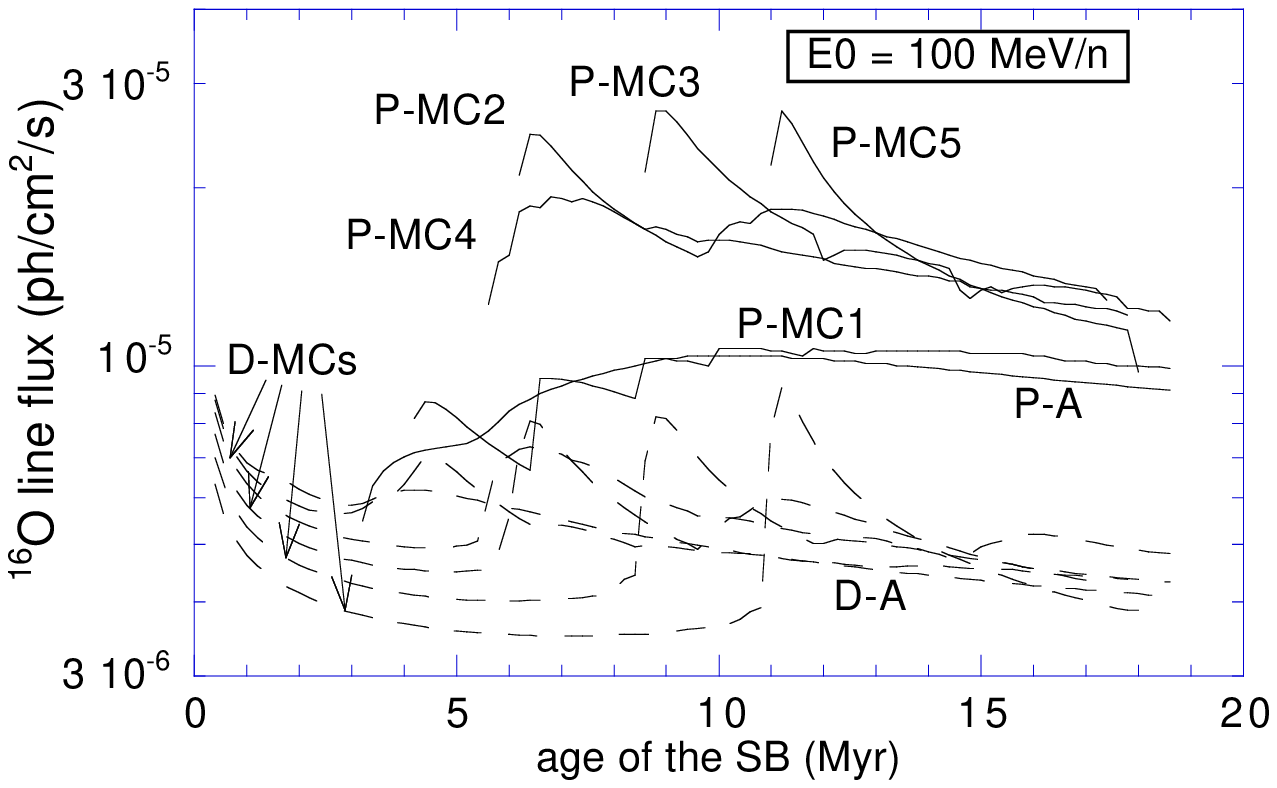, width=6.3cm, height=4cm}
            \hfill
            \psfig{file=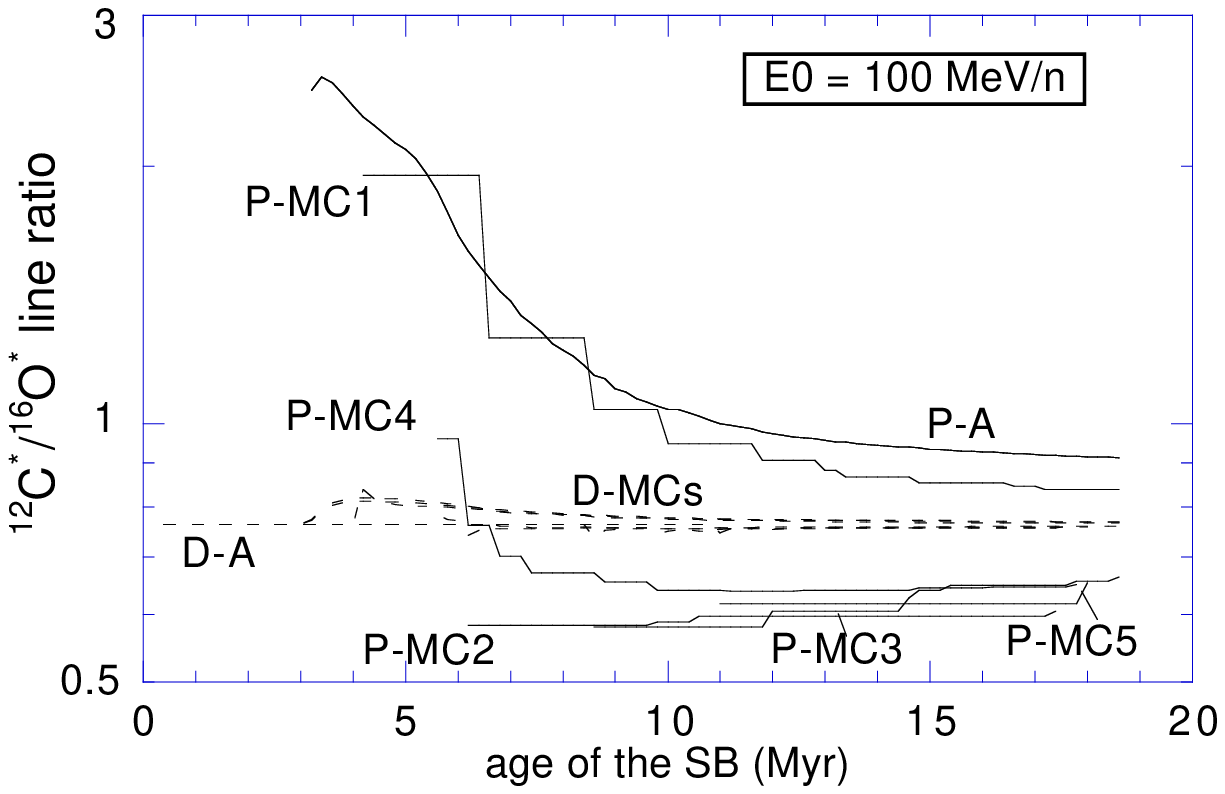, width=6.3cm, height=4cm}}
\caption{FIGURE 1. Orion-Eridanus superbubble evolution assuming
         $\Gamma=-1.7$ and 53 stars in the mass interval 4-15 $\Msol$
         (Brown et al.~1994).
         The leading letter indicates if pure (P) or
         fully mixed (D) ejecta were accelerated. Subsequent letters
         indicate analytic (A) or Monte Carlo simulated (MC)
         stellar populations.
         }
\end{figure}

The evolution of the Orion-Eridanus superbubble composition and the
predicted \gray\ line emission is summarised in Fig.~1.
On the one hand we populated the IMF using Monte Carlo samples that
are compatible with the present Orion population (Brown et al.~1994).
On the other hand we studied the academic case of an `analytic' stellar
population where the IMF is densely populated by `fractional' stars.
While the latter case provides the average \gray\ line emission, the
Monte Carlo sampling gives us the possible scatter around this average.

Our models predict 4.44 and 6.13 \MeV\ \gray\ line fluxes of the
order of a few $10^{-5}$~\funit, i.e.~around the expected threshold
of SPI for broadened lines (Jean 1996).
We want to emphasise that this value is only an order of magnitude
estimate due to the intrinsic uncertainties in our simplified model.
Nevertheless, the result indicates that Orion is still an interesting
target for the observation of \gray\ excitation lines due to its
proximity and star formation activity.
Our model predicts an \al\ production around $10^{-4}~\Msol$,
corresponding to 1.809 \MeV\ line fluxes of $\sim 6\,10^{-6}$~\funit.
This is compatible with the upper limit of COMPTEL (Oberlack et
al.~1995), and again is at the detection threshold of SPI.
In particular, the observation of either the 1.809 \MeV\ line or the
excitation lines (or both) will severely constrain the model
parameters and hence provide important information about shock
induced particle acceleration.

Among the most interesting observables is the
$^{12}\mathrm{C}^{\ast}/^{16}\mathrm{O}^{\ast}$ line ratio.  For
ejecta mixed with the evaporated ISM, the ratio is always very close
to the \emph{solar value} ($\sim~0.76$ for $E_{0} = 100$~\MeVn).  For
pure ejecta, this ratio may deviate significantly from the solar
value, with values depending on the presence of a very massive star
($M \ge 50 \Msol$) in the association (like in simulation MC1).
However, $^{12}\mathrm{C}^{\ast}/^{16}\mathrm{O}^{\ast}$ is also very
sensitive to the cut-off energy $E_{0}$ due to the different energy
dependencies of the excitation cross sections (cf.~Fig.~2).
Additionally, for $E_{0} < 20$ \MeVn\ the acceleration time scale
$\tau_{0}$ becomes too long, and hence the injection power too small,
for significant \gray\ line emission.  The ambiguity of interpreting a
given line ratio from the 3D-space of parameters ($E_{0}$, association
age, dilution) may be removed by jointly studying additional line
ratios, e.g.~$\mathrm{LiBe}^{\ast}/^{16}\mathrm{O}^{\ast}$ (where
LiBe$^{\ast}$ refers to the so-called Li-Be feature around 450~keV).
We will discuss the expected correlations in a separate paper where we
also give more detailed information about the modelling procedure
(Parizot \& Kn\"odlseder 1999, in preparation).

\begin{figure}
\centerline{\psfig{file=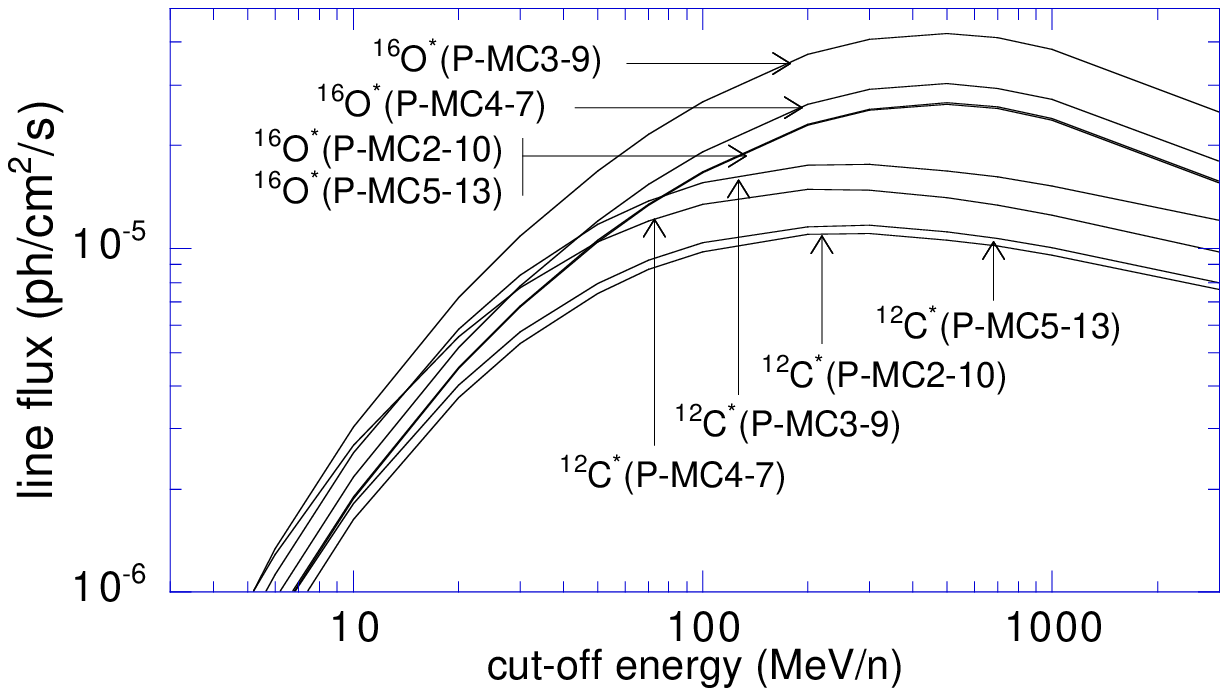, width=6.3cm, height=3.5cm}
            \hfill
            \psfig{file=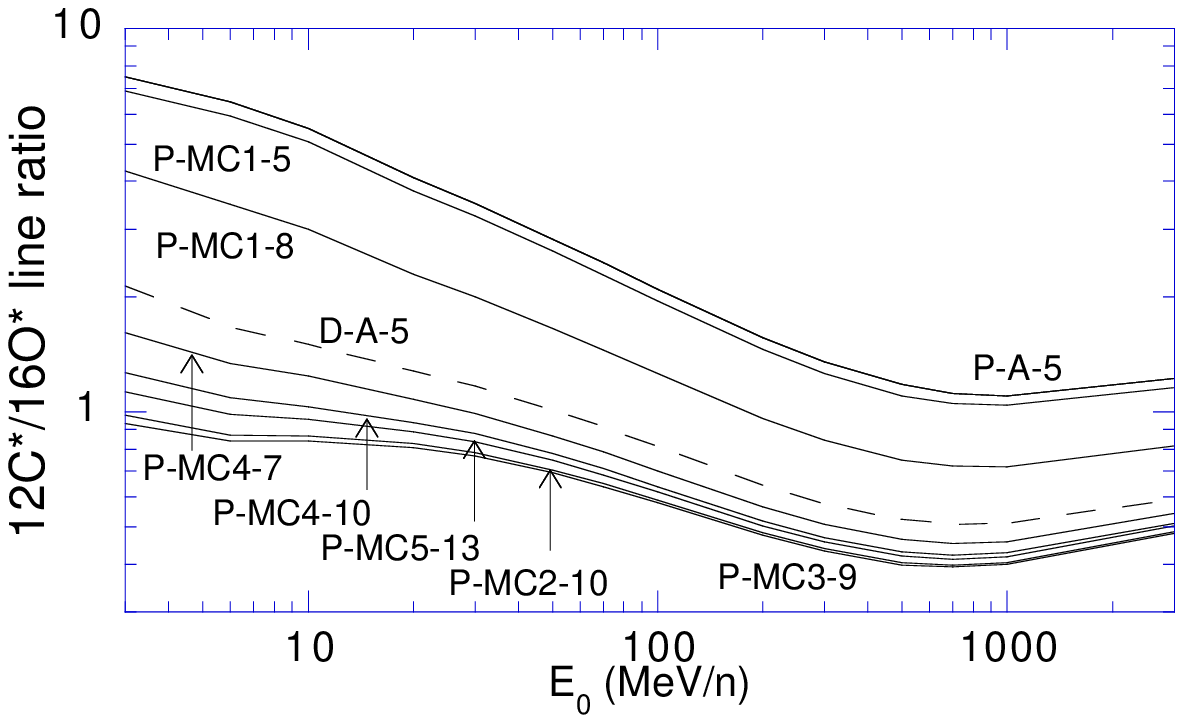, width=6.3cm, height=3.5cm}}
\caption{FIGURE 2. \gray\ line fluxes and
         $^{12}\mathrm{C}^{\ast}/^{16}\mathrm{O}^{\ast}$ line ratio as
         function of cut-off energy $E_{0}$ for different association
         ages $t_{6}$ ($\equiv 10^6$ years; labelled by an additional
         `-$t_{6}$').}
\end{figure}

}

%

\bsk
\baselineskip = 12pt

{\references \ni REFERENCES
\ssk

\ref Bloemen, H., et al.~1994, A\&A, 281, L5
\ref Brown, A.G.A., de Geus, E.J., \& de Zeeuw, P.T.~1994, A\&A, 289,
101
\ref Bykov, A.M.~\& Bloemen, H.~1994, A\&A, 283, L1
\ref Bykov, A.M.~\& Fleishman, G.D.~1992, MNRAS, 255, 269
\ref Bykov, A.M.~\& Toptygin, I.N.~1990, Sov.~Phys.-JETP, 71, 702
\ref Cass\'e, M., Lehoucq, R.~\& Vangioni-Flam, E.~1995, Nature, 373,
318
\ref Duncan, D., Lambert, D.~\& Lemke, M.~1992, ApJ, 401, 584
\ref Gilmore, G., et al.~1992, Nature, 357, 379
\ref Jean, P.~1996, PhD thesis, Univ.~Paul Sabatier, Toulouse
\ref Meynet, G., et al.~1997, A\&A, 320, 460
\ref Oberlack, U., et al.~1995, Proc.~ICRC, Rome, p.~207
\ref Parizot, E.~1998, A\&A, 331, 726
\ref Portinari, L., Chiosi, C., \& Bressan, A.~1998, A\&A, 334, 505
\ref Shull, J.M.~\& Saken, J.M.~1995, ApJ, 444, 663
\ref Woosley, S.E., Langer, N., \& Weaver, T.A.~1995, ApJ, 448, 315
\ref Woosley, S.E.~\& Weaver, T.A.~1995, ApJS, 101, 181
}

\end{document}